# Spin constrained orbital angular momentum control in high-harmonic generation

F. Kong[1,2], C. Zhang[1,2], H. Larocque[1], F. Bouchard[1], Z. Li[3], M. Taucer[1,2], G. Brown[1,2], T. J. Hammond[1,2,4], E. Karimi[1,5] & P. B. Corkum[1,2]

[1]*Department of Physics, University of Ottawa, 25 Templeton St., Ottawa K1N 6N5, Ontario, Canada*

[2] *Joint Attosecond Science Laboratory, University of Ottawa and National Research Council of Canada 100 Sussex Drive, Ottawa K1N 5A2, Ontario, Canada*

[3]*School of Optical and Electronic Information, Huazhong University of Science and Technology, Wuhan 430074, China*

[4]*Department of Physics, University of Windsor, Windsor N9B 3P4, Ontario, Canada*

[5]*Department of Physics, Institute for Advanced Studies in Basic Sciences, 45137-66731 Zanjan, Iran*

The interplay between spin and orbital angular momentum in the up-conversion process allows us to control the macroscopic wave front of high harmonics by manipulating the microscopic polarizations of the driving field. We demonstrate control of orbital angular momentum in high harmonic generation from both solid and gas phase targets using the selection rules of spin angular momentum. The gas phase harmonics extend the control of angular momentum to extreme-ultraviolet wavelength. We also propose a bi-color scheme to produce spectrally separated extreme-ultraviolet radiation carrying orbital angular momentum.

Structuring the spatial profiles of electro-magnetic fields can lead to important physics and applications [1,2]. Two physical quantities in particular are the phase front and the polarization, and have garnered great attention recently. For example, the wave front of Laguerre-Gaussian (LG) modes have a spiral shape, and is linked to the orbital angular momentum (OAM) of photons [3]. Such beams have been studied in high-speed communications, laser machining and quantum optics [4–6]. Due to the limitation of available optics [7,8], shaping the wave front of extreme-ultraviolet (XUV) or soft x-ray radiation is challenging, despite demand for these wavelengths in microscopy, spectroscopy and lithography applications [9–11]. Nonlinear frequency conversion under non-perturbative conditions provides a feasible route to control or transfer structured wave fronts to the XUV and even X-ray regions. By pre-shaping the driving optical field, the phase spirals that define the OAM can be imparted to the generated frequencies [12,13]. Similarly, the local polarization or the spin angular momentum (SAM) of the XUV field can also be controlled by engineering the polarization of the incident driving fundamental beam [14–16]. The circularly polarized XUV beams are a source for polarization sensitive measurements of inner shell electrons in materials and distinguishing chiral molecules [17–19].

Previous experiments show that, on their own, the spin and the orbital angular momentum are conserved during the nonlinear conversions in both perturbative [20–22] and non-perturbative regimes [12,23] under paraxial conditions. These experiments either modify the polarizations or the wave fronts of the driving laser beams, while keeping the other part uniform or planar. Such 'pure mode' configurations concentrate on studying the behavior of either the spin or the orbital angular momentum during the nonlinear conversion and deliberately eliminate possible influences from the other part of the angular momentum. So far, however, there has been little discussion about the interplay between these two types of angular momentum in high-order nonlinearities, which would allow us to further control the angular momentum in our desired wavelengths.

Instead of using a driving beam with a single spatial mode or a uniform polarization state, we manipulate both the spin and orbital angular momentum of the driving beam using one mode to control the other. In our specific case, the spin ($s=\pm 1$) selection rule eliminates the channels that could lead to higher OAM state [12,13], confining the OAM of the high harmonics to that of the fundamental, in our case $\pm 1$ unit. This interplay between the spin and orbital angular momentum during up-conversion allows us to control the macroscopic wave front in the short wavelength radiation by manipulating the microscopic polarization of the driving field.

The experiments are performed with both solid and gas targets. The solid target gives us access to both above and below bandgap harmonics, which can easily be manipulated and measured with conventional optics due to their low frequency. Harmonics generated from noble gases, on the other hand, allow this technique to be transferred to much higher frequencies. We will show that our results for photon energy ~42 eV are consistent with a beam with OAM values equal to ±1.

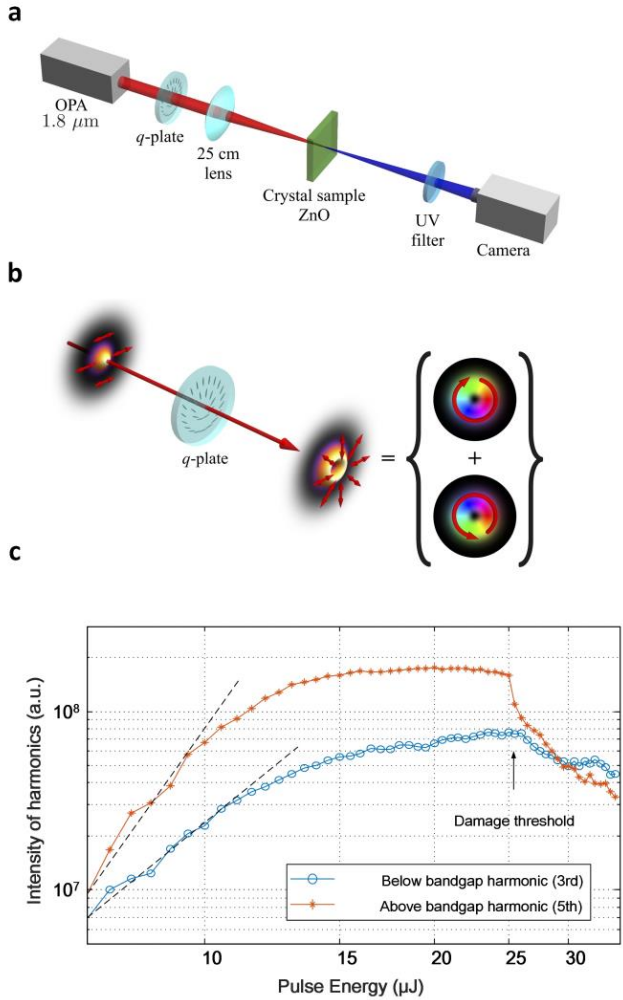

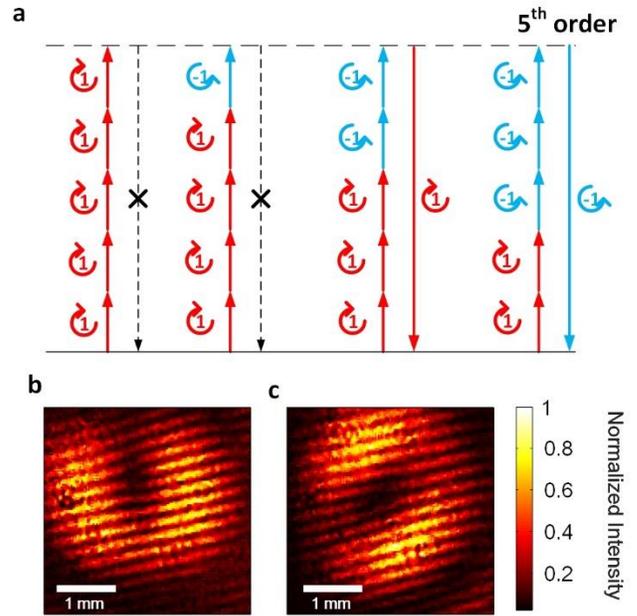

beam with a zero intensity at its center. The fundamental beam with ~20 μJ pulse energy is then focused by a 30 cm lens to reach ~1 TW/cm$^2$ on the crystal surface. Harmonics can be generated within either the perturbative or non-perturbative regimes before irreversible damage. The two limits are distinguished by intensity scaling (Fig. 1(c)) where both the 3$^{rd}$ and 5$^{th}$ harmonics saturate and no longer retaining its 3$^{rd}$ and 5$^{th}$ power law behavior (shown as the black dashed lines in Fig. 1(c)) [28]. The results presented below are obtained in the non-perturbative regime.

FIG. 1 Generating high-order harmonics under non-perturbative conditions. (a) The experimental setup of generating structured high-harmonic beams from a solid crystal target and interferometric characterization of the OAM on the harmonic beams. (b) A Gaussian beam is converted by a q-plate into a superposition of Laguerre Gaussian modes with opposite circular polarization states. (c) Intensity of scaling of harmonics yields with respect to different pulse energy of the driving lease pulse.

FIG. 2 Control of OAM on high-order harmonics by selection rules of SAM. (a) The allowed and forbidden transitions in multiphoton process plotted in energy diagram. (b) The experimentally observed interference pattern between a reference beam and a left handed circularly polarized OAM carrying beam with $l=+1$. (c) The experimentally observed interference pattern between a reference beam and a right handed circularly polarized OAM carrying beam with $l=-1$.

In our experiment a laser beam (with a duration of 50 fs, center wavelength of 1.8 μm for ZnO crystal and 800 nm for Argon gas) generates harmonics in a solid or gas-phase target (Fig. 1(a)). We shape the spatial characteristics of the beam with a *q*-plate – a liquid-crystal device that can change the polarization (or phase) of the incident laser beam [24–26] point by point across the beam profile. After passing through the *q*-plate, our 1.8 μm beam in LG$_{0,0}$ mode is converted to a superposition of LG$_{0,1}$ and LG$_{0,-1}$ modes [27], as illustrated in Fig. 1(b). These two modes have opposite circular polarization, corresponding to two eigenstates of SAM, and their superposition results in a radially polarized

During harmonic generation, energy conservation requires that the number of fundamental photons involved in producing harmonics must equal the harmonic order. For instance, in 5$^{th}$ order harmonic generation shown in Fig. 2(a), five fundamental photons are absorbed and produce one ultraviolet photon emitted at 360 nm. This emitted ultraviolet (UV) photon can only have two possible spin eigenstates, $s=+1$ and $s=-1$, where $s$ is the quantum number of SAM. Therefore, only two possible channels are allowed: either absorbing two left ($s=+1$) and three right circularly polarized photons ($s=-1$), or three left ($s=+1$) and two right ($s=-1$), as shown in Fig. 2(a) on the right. Any other combinations, for example those shown on the left, are forbidden, since the final SAM of the UV photon cannot satisfy the condition $\langle s \rangle \leq 1$ when spin is conserved.

The total OAM of the emitted harmonic photons for the two allowed processes equals to the sum of OAM from all involved fundamental photons. In our case, the OAM value for the two processes can only be $l=+1$ and $l=-1$, respectively, where $l$ is the quantum number of the OAM states. In other words, the OAM of the harmonics equals the OAM of the fundamental. It is the bounded spin states and conservation of angular momentum which selects the OAM value of harmonics.

To show that the OAM value of each circularly polarized state is constrained by the spin selection rule, we use a combination of a quarter-wave plate and a linear polarizer to serve as an analyzing element for circular polarization [29]. Since these two optical components do not change the topology of the wave front, the OAM quantum number can be measured by interference with another plane wave sharing the same polarization state [30]. The reference plane wave is coherently created by an independent source from the solid target without wave front shaping.

Fig. 2(b,c) shows the interferometric results of the $s=+1$ and $s=-1$ state 5th harmonics – above-bandgap harmonics for ZnO. Both results show a fork-shaped pattern and opposite orientations between two circular polarized states. The beam's OAM value can be read from different numbers of fringes from left side of singularity point to its right side. The OAM values on the $s=\pm 1$ components are $l=\pm 1$, respectively, which correspond to the OAM of the fundamental beam.

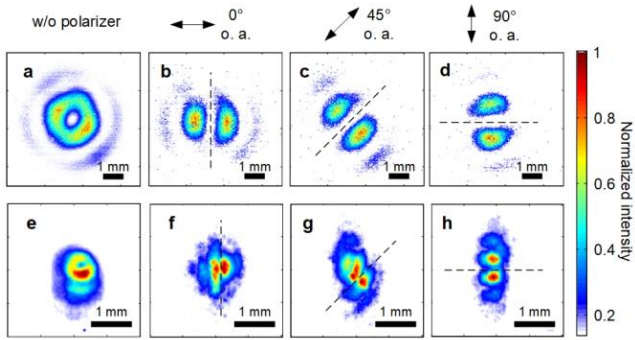

FIG. 3 Characterization of polarization states of high-harmonic beams using a linear polarizer. (a) Intensity profile of generated 5th harmonics from ZnO target. (b-d) The intensity profiles of 5th harmonic beams from ZnO, with wavelength of 360 nm, after passing through a linear polarizer with its optical axis angle placed at 0°, 45° and 90°, respectively. (e) Intensity profile of generated high-order harmonics from argon gas target. (f-h) The intensity profiles of 27th harmonic beams from argon gas, with wavelength of 30 nm, after passing through a linear polarizer with its optical axis angle placed at 0°, 45° and 90°, respectively.

It is possible for the SAM of the optical beam to be transferred to the crystal lattice [31,32], depending on the rotational symmetry of the crystal sample. In our experiment with (0001)-cut free-standing ZnO, such an effect is not excluded. However, the measured harmonic yields from these channels are below the noise level in our measurement, which are at least two orders-of-magnitude weaker than the channels shown in Fig. 2(a) that do not exchange SAM with the crystal lattice. Therefore, the higher order OAM states are negligible, and the $l=\pm 1$ state dominates the harmonic output.

Symmetry-enabled channels that lead to higher-order OAM modes may be revealed in other strongly coupled systems. For such processes to be important, the material should exhibit an anisotropic harmonic yield with varying orientation of a linearly polarized driving field [16,33].

As a superposition of two equally intense circular states, the driving field is locally linear everywhere at its focus. Thus, the generated harmonic beams should maintain the radially polarized structure of the driving beam. The polarization singularity forces a zero intensity at the beam center as shown in Fig. 3(a). To confirm the polarization of the generated harmonic beam, we measure the polarization of the 5th harmonic beam using a linear polarizer, a UV band-pass filter and a UV enhanced camera. The intensity profiles of the 5th harmonic after passing through the linear polarizer is shown in Fig. 3(b-d) where we have oriented the linear polarizer at 0°, 45° and 90°, respectively. The bright lobes rotate as we change the angle of the linear polarizer. The horizontal/vertical parts of the beam are observed when the optical axis of the linear polarizer is placed horizontally/vertically. This is consistent with the characteristic of a radially polarized beam. From an interferometric point of view, in circular-state bases, the linear polarizer selects a common linear component from the two circular states and lets them interfere collinearly on the camera. The two bright lobes in their intensity illustrate $2 \times 2\pi$ phase shift between left and right circular states in the azimuthal direction. This confirm the analysis that one unit of OAM is imparted to the 5th harmonics with opposite signs between the two spin states.

To reach higher photon energies, we extend the above experiment to a gas-phase target and a much higher intensity. With a noble gas, the effect of transferring SAM to the nonlinear medium is also eliminated, due to the rotational symmetry of the gas atoms [32]. We use the optical pulse at 800 nm with 50 fs duration and 0.8 mJ energy to produce high harmonics from argon gas. The emission extends to the 29th harmonics of 800nm with photon energy of ~45 eV. The harmonic emission from the gas reproduces a doughnut-shaped intensity distribution from the driving beam, which is shown in Fig. 3(e). The image is recorded directly by a micro-channel plate without passing through dispersive element, to eliminate the polarization selectivity from the measuring system. Therefore, the intensity distribution is an intensity summation of all high harmonics.

However, it is not trivial to separate the two superimposed circularly polarized beam at XUV wavelength and measure their OAM value separately. To link the result from gas and solid targets, we replace the linear polarizer used for solids by a pair of silver mirrors [34]. The generated high harmonics are passed through both the mirror pair and the XUV imaging grating before detected by the micro-channel plate. The polarization selectivity of the mirror pair and the grating is ~10:1. Instead of rotating the mirror pairs, we rotate the incident beam to change the optical axis of the linear polarizer. As shown in Fig. 3(f-h), we see two nodes orient along different angles as we rotate the driving laser beam. This is the same behavior as we reported in Fig. 3(a-c) for a solid target, and it indicates that controllable OAM is also transferred to XUV wavelength.

To decouple two superimposed circular states at XUV wavelength, we propose a bi-color driving approach to isolate high harmonics beam with controlled OAM. In that case, the driving field would consist of a fundamental OAM beam with $l$=+1, $s$=+1 and a second harmonic beam with $l$=-1, $s$=-1. According to the conservation of SAM, there are only two allowed channels for high harmonic generation, as shown in Fig. 4. Different from the scheme driven by a single-color field, the emissions from these two channels corresponds to $7^{th}$ and $8^{th}$ harmonic generation. In other words, the two OAM modes are decoupled in energy. Classically, the mixing of the two-color field will result in a 3-fold symmetric driving field, with a re-collision happening every third of a period [35]. Since the two beams carry different phase front spiral, the 3-fold trajectories rotate their orientations by $2\pi$ along azimuthal direction. This geometric phase gives rise to the phase-front spiral of the generated high harmonics.

Controllable OAM (or structured polarization states) of short wavelength radiation will result in tighter optical focusing [36], excite inner shell dynamics [37] and probe ultrafast magnetic dynamics [38]. Our study contributes to this growing area of research by exploring the relation between the spin and orbital angular momentum in the up-conversion process under strong field condition. Furthermore, we show the potential of solving control problems by linking the two components of the angular momentum. The topological charge is constrained to be equal to that of the fundamental beam. It can, therefore, be easily manipulated by the conventional optical elements [12,13]. In addition, the collinear geometry and locally linearly polarized driving field ensures efficient conversion to the harmonic emission, which is crucial for the development of a light source.

In our experiment, both spin and orbital angular momentum are conserved on their own, since the medium that we are interacting with is nearly isotropic and the focusing geometry of the driving laser beam is paraxial. However, exploring the interaction, for example in waveguide or tightly focused geometry [39,40] where spin and orbital angular momentum can be coupled, may lead to even more complex behavior. Just like other fundamental particles, photons exhibit spin-orbit coupling when interacting with matter. The coupling not only exists between spin and orbital angular momentum, as we mentioned above, it can also happen between the photons and crystal/electronic structures. A strongly coupled solid-phase system may help create electro-optic devices that respond to the polarization and wave front of incident light, or new optical sources, such as polarization/spin controlled soft x-ray for magnetically sensitive probing.

The authors are grateful to Professor Robert W. Boyd and Professor Jeff Lundeen from University of Ottawa for their expert advice and helpful criticism during the elaboration of this work. We are also pleased to acknowledge the support of Dr. Andrei Naumov and David Crane from National Research Council of Canada and Hsuan Wang from University of Ottawa throughout this work. Finally, we acknowledge financial support from the Canada Research Chair (CRC) and Canada Foundation for Innovation (CFI) and the US DARPA grant under the TEE program

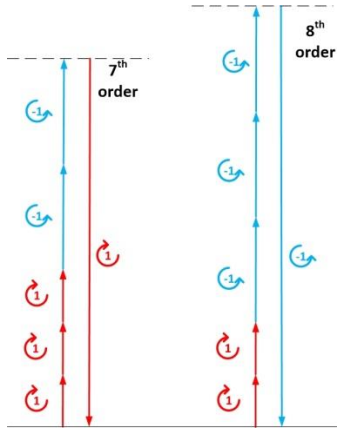

FIG. 4 The energy diagram of spectrally decoupled OAM mode in high-order harmonics. Two allowed 5-photon processes in the two-color scheme is split into $7^{th}$ and $8^{th}$ harmonics in spectrum, whereas the emitted photons are energy degenerate using the same color driving field.

[1] Q. Zhan, Adv. Opt. Photonics **1**, 1 (2009).
[2] G. Molina-Terriza, J. P. Torres, and L. Torner, Nat. Phys. **3**, 305 (2007).
[3] L. Allen, M. W. Beijersbergen, R. J. C. Spreeuw, and J. P. Woerdman, Phys. Rev. A **45**, 8185 (1992).
[4] A. Mair, A. Vaziri, G. Weihs, and A. Zeilinger, Nature **412**, 313 (2001).
[5] J. Wang, J.-Y. Yang, I. M. Fazal, N. Ahmed, Y. Yan, H. Huang, Y. Ren, Y. Yue, S. Dolinar, M. Tur, and A. E. Willner, Nat. Photonics **6**, 488 (2012).

[6] K. Toyoda, F. Takahashi, S. Takizawa, Y. Tokizane, K. Miyamoto, R. Morita, and T. Omatsu, Phys. Rev. Lett. **110**, 143603 (2013).
[7] W. Harm, S. Bernet, M. Ritsch-Marte, I. Harder, and N. Lindlein, Opt. Express **23**, 413 (2015).
[8] L. Shi, Z. Zhang, A. Cao, X. Luo, and Q. Deng, Opt. Express **23**, 8620 (2015).
[9] S. Fuchs, C. Rödel, A. Blinne, U. Zastrau, M. Wünsche, V. Hilbert, L. Glaser, J. Viefhaus, E. Frumker, P. Corkum, E. Förster, and G. G. Paulus, Sci. Rep. **6**, 20658 (2016).
[10] H. J. Wörner, J. B. Bertrand, D. V. Kartashov, P. B. Corkum, and D. M. Villeneuve, Nature **466**, 604 (2010).
[11] C. Wagner and N. Harned, Nat. Photonics **4**, 24 (2010).
[12] G. Gariepy, J. Leach, K. T. Kim, T. J. Hammond, E. Frumker, R. W. Boyd, and P. B. Corkum, Phys. Rev. Lett. **113**, 153901 (2014).
[13] F. Kong, C. Zhang, F. Bouchard, Z. Li, G. G. Brown, D. H. Ko, T. J. Hammond, L. Arissian, R. W. Boyd, E. Karimi, and P. B. Corkum, Nat. Commun. **8**, 14970 (2017).
[14] O. Kfir, P. Grychtol, E. Turgut, R. Knut, D. Zusin, D. Popmintchev, T. Popmintchev, H. Nembach, J. M. Shaw, A. Fleischer, H. Kapteyn, M. Murnane, and O. Cohen, Nat. Photonics **9**, 99 (2014).
[15] P. Antoine, A. L'Huillier, M. Lewenstein, P. Salières, and B. Carré, Phys. Rev. A **53**, 1725 (1996).
[16] N. Saito, P. Xia, F. Lu, T. Kanai, J. Itatani, and N. Ishii, Optica **4**, 1333 (2017).
[17] A. Ferré, C. Handschin, M. Dumergue, F. Burgy, A. Comby, D. Descamps, B. Fabre, G. A. Garcia, R. Géneaux, L. Merceron, E. Mével, L. Nahon, S. Petit, B. Pons, D. Staedter, S. Weber, T. Ruchon, V. Blanchet, and Y. Mairesse, Nat. Photonics **9**, 93 (2015).
[18] T. Fan, P. Grychtol, R. Knut, C. Hernández-García, D. D. Hickstein, D. Zusin, C. Gentry, F. J. Dollar, C. A. Mancuso, C. W. Hogle, O. Kfir, D. Legut, K. Carva, J. L. Ellis, K. M. Dorney, C. Chen, O. G. Shpyrko, E. E. Fullerton, O. Cohen, P. M. Oppeneer, D. B. Milošević, A. Becker, A. A. Jaroń-Becker, T. Popmintchev, M. M. Murnane, and H. C. Kapteyn, Proc. Natl. Acad. Sci. U. S. A. **112**, 14206 (2015).
[19] G. Lambert, B. Vodungbo, J. Gautier, B. Mahieu, V. Malka, S. Sebban, P. Zeitoun, J. Luning, J. Perron, A. Andreev, S. Stremoukhov, F. Ardana-Lamas, A. Dax, C. P. Hauri, A. Sardinha, and M. Fajardo, Nat. Commun. **6**, 6167 (2015).
[20] G. Walker, A. S. Arnold, and S. Franke-Arnold, Phys. Rev. Lett. **108**, 243601 (2012).
[21] N. V. Bloch, K. Shemer, A. Shapira, R. Shiloh, I. Juwiler, and A. Arie, Phys. Rev. Lett. **108**, 233902 (2012).
[22] K. Dholakia, N. B. Simpson, M. J. Padgett, and L. Allen, Phys. Rev. A **54**, R3742 (1996).
[23] A. Fleischer, O. Kfir, T. Diskin, P. Sidorenko, and O. Cohen, Nat. Photonics **8**, 543 (2014).
[24] H. Larocque, J. Gagnon-Bischoff, F. Bouchard, R. Fickler, J. Upham, R. W. Boyd, and E. Karimi, J. Opt. **18**, 124002 (2016).
[25] F. Cardano, E. Karimi, S. Slussarenko, L. Marrucci, C. de Lisio, and E. Santamato, Appl. Opt. **51**, C1 (2012).
[26] S. Slussarenko, A. Murauski, T. Du, V. Chigrinov, L. Marrucci, and E. Santamato, Opt. Express **19**, 4085 (2011).
[27] E. Karimi, B. Piccirillo, L. Marrucci, and E. Santamato, Opt. Lett. **34**, 1225 (2009).
[28] M. Sivis, M. Taucer, G. Vampa, K. Johnston, A. Staudte, A. Y. Naumov, D. M. Villeneuve, C. Ropers, and P. B. Corkum, Science **357**, 303 (2017).
[29] J. Leach, J. Courtial, K. Skeldon, S. M. Barnett, S. Franke-Arnold, and M. J. Padgett, Phys. Rev. Lett. **92**, 13601 (2004).
[30] M. Harris, C. A. Hill, P. R. Tapster, and J. M. Vaughan, Phys. Rev. A **49**, 3119 (1994).
[31] H. J. Simon and N. Bloembergen, Phys. Rev. **171**, 1104 (1968).
[32] C. L. Tang and H. Rabin, Phys. Rev. B **3**, 4025 (1971).
[33] Y. S. You, D. A. Reis, and S. Ghimire, Nat. Phys. **13**, 345 (2017).
[34] J. Levesque, Y. Mairesse, N. Dudovich, H. Pépin, J.-C. Kieffer, P. B. Corkum, and D. M. Villeneuve, Phys. Rev. Lett. **99**, 243001 (2007).
[35] A. Fleischer, O. Kfir, T. Diskin, P. Sidorenko, and O. Cohen, Nat. Photonics **8**, 543 (2014).
[36] R. Dorn, S. Quabis, and G. Leuchs, Phys. Rev. Lett. **91**, 233901 (2003).
[37] M. Uiberacker, T. Uphues, M. Schultze, A. J. Verhoef, V. Yakovlev, M. F. Kling, J. Rauschenberger, N. M. Kabachnik, H. Schröder, M. Lezius, K. L. Kompa, H.-G. Muller, M. J. J. Vrakking, S. Hendel, U. Kleineberg, U. Heinzmann, M. Drescher, and F. Krausz, Nature **446**, 627 (2007).
[38] A. V. Kimel, A. Kirilyuk, P. A. Usachev, R. V. Pisarev, A. M. Balbashov, and T. Rasing, Nature **435**, 655 (2005).
[39] K. Y. Bliokh, E. A. Ostrovskaya, M. A. Alonso, O. G. Rodríguez-Herrera, D. Lara, and C. Dainty, Opt. Express **19**, 26132 (2011).
[40] K. Y. Bliokh, F. J. Rodríguez-Fortuño, F. Nori, and A. V. Zayats, Nat. Photonics **9**, 796 (2015).